\documentclass[10pt,showpacs,preprintnumbers,amsmath,amssymb,
a4paper,nofootinbib]{revtex4}

\usepackage{amsmath,amssymb,graphicx}
\usepackage{epsf}
\usepackage{epstopdf}
\usepackage {amssymb}

\def\beq{\begin{equation}}
\def\eeq{\end{equation}}
\def\bea{\begin{eqnarray}}
\def\eea{\end{eqnarray}}

\newcommand{\nc}{\newcommand}
\nc{\ba}{\begin{eqnarray}}
\nc{\ea}{\end{eqnarray}}

\newcommand\eq[1]{Eq.~(\ref{#1})}

\begin{document}

\title{The Effect of Curvaton Decay on the Primordial Power Spectrum}

\author{Hassan Firouzjahi$^{1}$}
\email{firouz-AT-mail.ipm.ir}
\author{Anne Green$^{2}$}
\email{anne.green(AT)nottingham.ac.uk}
\author{Karim Malik$^{3}$}
\email{k.malik(AT)qmul.ac.uk}
\author{Moslem Zarei$^{4,5}$}
\email{m.zarei(AT)cc.iut.ac.ir}
\affiliation{$^1$School of Astronomy, Institute for Research in 
Fundamental Sciences (IPM),
P. O. Box 19395-5531,
Tehran, Iran}
\affiliation{$^2$School of Physics and Astronomy, University of 
  Nottingham, University Park, Nottingham, NG7 2RD, UK}
\affiliation{$^3$Astronomy Unit, School of Physics and Astronomy, Queen Mary University of London,
Mile End Road, London, E1 4NS, United Kingdom}
\affiliation{$^4$Department of Physics, Isfahan University of Technology, Isfahan 84156-83111, Iran }
\affiliation{$^5$School of Physics, Institute for Research in 
Fundamental Sciences (IPM), P. O. Box 19395-5531, Tehran, Iran}

\date{\today}

\begin{abstract}
We study the effect of curvaton decay on the primordial power
spectrum. Using analytical approximations and also numerical
calculations, we find that the power spectrum is enhanced during the
radiation dominated era after the curvaton decay. The amplitude of the
Bardeen potential is controlled by the fraction of the energy density
in the curvaton at the time of curvaton decay.  We show that the
enhancement in the amplitude of the primordial curvature perturbation
is, however, not large enough to lead to primordial black hole
overproduction on scales which re-enter the horizon after the time of
curvaton decay.
\end{abstract}

\pacs{98.80.Cq \hfill  arXiv:1209.xxxx}

\maketitle


\section{Introduction}

Recent results from experiments at the LHC in Geneva give a very
strong indication of the existence of a Higgs like particle. If
confirmed, this would be the first observation of a scalar field
outside cosmology. For many years scalar fields have been nearly
indispensable in theoretical cosmology, not only in solving many
problems of the hot big bang model, but also in generating the
primordial power spectrum of density fluctuations (see
e.g. Ref.~\cite{LL}).

The curvaton scenario \cite{Lyth:2001nq,Enqvist:2001zp,Moroi:2001ct}
is an elegant extension to the standard inflation scenario. Whereas in
the standard scenario the fields driving inflation and the field
generating the primordial power spectrum are the same, in the curvaton
scenario these tasks are performed by two (or more) fields. While the
inflaton field drives inflation, the curvaton field is a mere
spectator field (massless or nearly massless), picking up a nearly
scale invariant spectrum during inflation. After inflation the
inflaton field decays into radiation and the curvaton behaves like
dust, oscillating around the minimum of its potential. After this
brief matter dominated period the curvaton also decays into
radiation. Due to its simplicity and elegance the curvaton scenario
has enjoyed widespread attention in recent years
(see e.g.~Refs.~\cite{Bartolo:2002vf,Lyth:2002my,Lyth:2003ip,Dimopoulos:2003ss,Bartolo:2003jx,Malik:2006pm,Sasaki:2006kq, Assadullahi:2007uw, Enqvist:2008be,Dimopoulos:2012nj,Assadullahi:2012yi}.

In most standard curvaton calculations the power spectrum of the
curvature perturbation $\zeta$ (the quantity that later on sources the
 CMB and LSS) is assumed to be directly inherited from the curvaton
field. In this paper we study in detail the effect of curvaton decay
on the primordial power spectrum. In particular we investigate whether
the change in the initial power spectrum on small scales can lead to
primordial black hole (PBH) production. To keep our results as general
as possible, we do not specify a potential for the curvaton field,
assuming only that it oscillates around the minimum of its potential,
and hence behaves like dust, before decaying into radiation.\\

The outline of the paper is as follows. In Sections~\ref{bd_sect} and
\ref{pert_sect} we present the equations governing the background and
perturbation evolution. In Section \ref{bardeen_sect} we derive
analytical approximations for the Bardeen potential, and calculate the
effect of curvaton decay on the power spectrum of $\zeta$. We then
calculate the abundance of PBHs formed in Section \ref{pbh_sect}. We
conclude with a brief discussion in the final section.

\section{Background Dynamics }
\label{bd_sect}

In this section we provide the background dynamics. We consider two interacting fluids in a
Friedmann-Robertson-Walker (FRW) universe: radiation with equation of
state $P\equiv \omega_{\rm r}\rho_{\rm r}$, where $\omega_{\rm r}=1/3$ and
the curvaton with dust like equation of state. 
We assume the inflaton entirely decays into radiation at the end of inflation. Initially radiation dominates and the curvaton 
energy density is very sub-dominant. We assume that the curvaton
decays into radiation with a constant decay rate
$\Gamma$ when the Hubble expansion rate $H$ drops below $\Gamma$.


\subsection{Background Equations}


The continuity equations for each individual fluid in the background
can be written as \cite{wands1,wands2}
\bea
 && \dot{\rho}_{\sigma}+3H\rho_{\sigma}=-\Gamma\rho_{\sigma}  \,, \label{densityeq1} \\ &&
\dot{\rho}_{r}+4H\rho_{r}=\Gamma\rho_{\sigma} \label{densityeq2} \, ,
 \eea
where $\rho_{\sigma}$ and $\rho_{\rm r}$ are the background energy density
of the curvaton and radiation respectively and a dot indicates the derivative with respect to cosmic time $t$.
The right-hand side of
equations (\ref{densityeq1}) and (\ref{densityeq2}) describe the
background energy transfer per unit time from the curvaton and to the
radiation fluid respectively. The Hubble parameter $H$ is determined by the
Friedmann constraint
\beq
 H^{2}=\frac{1}{3\,M^{2}_{\rm{Pl}}}\:(\rho_{\sigma}+\rho_{r}) \, ,
\eeq
where $M_{\rm{Pl}}\equiv 1/\sqrt{8\pi G}$ is the reduced Planck mass. In order to obtain the
evolution of the above dynamical system, it is useful to write the
equations in terms of dimensionless parameters
\beq
\Omega_{\sigma}=\frac{\rho_{\sigma}}{\rho},\:\:\:\:\:\: \textrm{and}\:\:\:\:\: \Omega_{r}=\frac{\rho_{r}}{\rho} \, ,
\eeq
where
\beq
\sum_{\alpha}\rho_{\alpha}=\rho \,,
\eeq
is the total energy density and hence $\Omega_\sigma + \Omega_r =1$.
One can easily derive
the following evolution equations for these dimensionless parameters:
\bea
\label{back1}
&& \Omega^{'}_{\,\sigma}(N)-\Omega_{\,\sigma}(N)\Omega_{\,r}(N)+\frac{\Gamma}{H(N)}\,\Omega_{\,\sigma}(N)=0 \label{back7} \,,\\  &&
\label{back2}
\Omega^{'}_{\,r}(N)+ \Omega_{\,r}(N) \Omega_{\sigma}(N)-\frac{\Gamma}{H(N)}\,\Omega_{\,\sigma}(N)=0 \label{back8} \,,\\  &&
\label{back3}
H^{'}(N)+\frac{3}{2}H(N)+\frac{1}{2}H(N)\Omega_{\,r}(N)=0 \label{back9} \, ,
\eea
where a prime denotes differentiation with respect to the number of
e-folding $N\equiv\int H\textrm{d}t$. We set $N=0$ at the end of reheating.

\subsection{Sudden decay limit}

Useful intuition can be gained by first considering the analytical solutions of the system of Eqs.~(\ref{back1}--\ref{back3}) in  the sudden decay limit 
 where $\Gamma =0$.  This is a good approximation until the time of
curvaton decay at $N=N_{\rm d}$ when $\Gamma \sim H$ and the curvaton decays to radiation. In the sudden decay limit where $\Gamma/H=0$, Eqs.~(\ref{back1}), (\ref{back2}) and
(\ref{back3}) can be solved easily for the background quantities $\Omega_{\rm r}(N), \Omega_\sigma(N)$, and $H(N)$, to obtain
\ba
\label{sol-1}
\Omega_{\rm r}(N)  &=& \frac{1}{1+ p \, e^{N}} \,, \label{solback1} \\
\Omega_\sigma(N)  &=& \frac{p \, e^{N}}{1+ p \, e^{N}} \,,   \label{solback1b}\\
H(N)&=& H_0 e^{-2 N} \left( 1+ p \, e^N \right)^{1/2} \,, \label{solbackH} \quad
\ea
where $H_0$ is the value of $H$ at the start of radiation dominated epoch, after the end of reheating, when $N=0$. Here we have defined the ratio $p$
\ba
{ p} \equiv \frac{\Omega_{\sigma \, {\rm in}}}{\Omega_{{\rm r}\, {\rm in}}} \,,
\ea
where $\Omega_{\sigma \, {\rm in}}$ and $\Omega_{{\rm r} \, {\rm in}}$ are the initial values of
$\Omega_{\sigma }$ and $\Omega_{\rm r}$ at the end of reheating, so that if radiation initially dominates $p \ll 1$.

A key parameter in curvaton analysis is $f_{\rm d}$, the weighted fraction of curvaton energy density to the total energy density at the time of curvaton decay, defined by
\ba
\label{fd-def}
f_{\rm d} \equiv \frac{3 \rho_{\sigma}}{3 \rho_{\sigma} + 4 \rho_{r}}{\big|}_{\mathrm{dec}} 
= \frac{3 \Omega_\sigma}{4- \Omega_\sigma}{\big|}_{\mathrm{dec}}
\sim  \Omega_\sigma|_{\mathrm{dec}} \, .
\ea
As is well-known \cite{Lyth:2001nq, Lyth:2002my}, the non-Gaussianity parameter $f_{\rm NL}$ on CMB scales is approximately given by
$f_{\rm NL} \sim 1/f_{\rm d}$.
In Section \ref{bardeen_sect} we find the Bardeen potential $\Psi$ for various limiting values of $f_{\rm d}$.

\subsection{Beyond the sudden decay limit}

Here we provide solutions for $\Omega_{\rm r}(N)$ and
$\Omega_\sigma(N)$ which can also be used after the time of curvaton
decay. Comparison with the full numerical solution to
Eqs.~(\ref{back1}--\ref{back3}) shows that while our analytical
solution for $H(N)$ in the sudden decay limit, Eq.~(\ref{solbackH}) is
in reasonable agreement with the numerical solution the analytical
sudden decay solutions for $\Omega_{\rm r}(N)$ and $
\Omega_\sigma(N)$, Eqs.~(\ref{solback1}) and (\ref{solback1b}), are
not accurate for $N>N_{\rm d}$ as expected. This indicates that more
accurate analytical solutions for $\Omega_{\rm r}(N)$ and $
\Omega_\sigma(N)$ can be found by inserting the solution for $H(N)$
into Eqs.~(\ref{back1}) and (\ref{back2}) and then solving for
$\Omega_{\rm r}(N)$ and $ \Omega_\sigma(N)$.

Defining
\ba
X \equiv \frac{\Omega_\sigma}{\Omega_r} \, ,
\ea
from Eqs. (\ref{back1}) and  (\ref{back2})  we find
\ba
\label{X-eq}
X' = X -\frac{\Gamma}{H} (1+ X) X\, ,
\ea
with $H(N)$ given by Eq.~(\ref{solbackH}). It is not possible to solve
Eq.~(\ref{X-eq}) in general. However, away from the time of curvaton decay, we have
$X<1$, i.e. $\Omega_\sigma < \Omega_{\rm r}$ so Eq.~(\ref{X-eq}) can be approximated by 
a linear differential equation which can be solved analytically yielding
\ba
\label{X-sol}
X  &\simeq& p \exp { \left[ \int_0^N {\rm d} N' \left(1- \frac{\Gamma}{H(N')} \right)     \right]} \,, \\
\label{X-sol2}
&\simeq & p \exp\left\{ -\frac{2 \Gamma}{H_0 p^2} \left[ \sqrt{1+ p e^{N} } \left(\frac{1+ p e^N}{3}-1\right) + \frac{2}{3}\right] +N \right\} \, .
\ea
In the integrand above $H(N')$ is calculated from the sudden decay limit solution given in Eq.~(\ref{solbackH}). Correspondingly
\ba
\label{sol-sigma}
\Omega_r(N)  = \frac{1}{1+X} \quad , \quad \Omega_\sigma(N) = \frac{X}{1+X} \, .
\ea
Note that before the time of curvaton decay when one can neglect the last term
in Eq.~(\ref{X-eq}) and $X \simeq p e^{N}$,  Eq.~(\ref{sol-sigma}) reduces to
Eqs.~(\ref{solback1}) and (\ref{solback1b}) obtained in the sudden decay limit.

For the physically interesting case in which the curvaton makes a
sub-dominant contribution to the total energy density at the time of
decay, corresponding to $\Omega_\sigma|_{\rm dec} \ll 1$, one can take
$\Omega_\sigma(N) \ll \Omega_{\rm r}(N)$ throughout the whole
evolution and to a good approximation $H(N)$ is given, as in a
radiation dominated Universe, by $H(N)\simeq H_0 e^{2N}$. Inserting
this expression for $H$ into Eq.~(\ref{X-sol}) yields
\ba
\label{limit}
\Omega_\sigma \simeq  p \exp\left(N- \frac{\Gamma}{2 H_0} e^{2N}   \right) \,, \quad
\Omega_r  = 1- \Omega_{\sigma} \,,
\quad \quad \quad \quad       \left( f_{\rm d}\ll 1 \right)\,.
\ea

\begin{figure}
\includegraphics[width=0.4\linewidth]{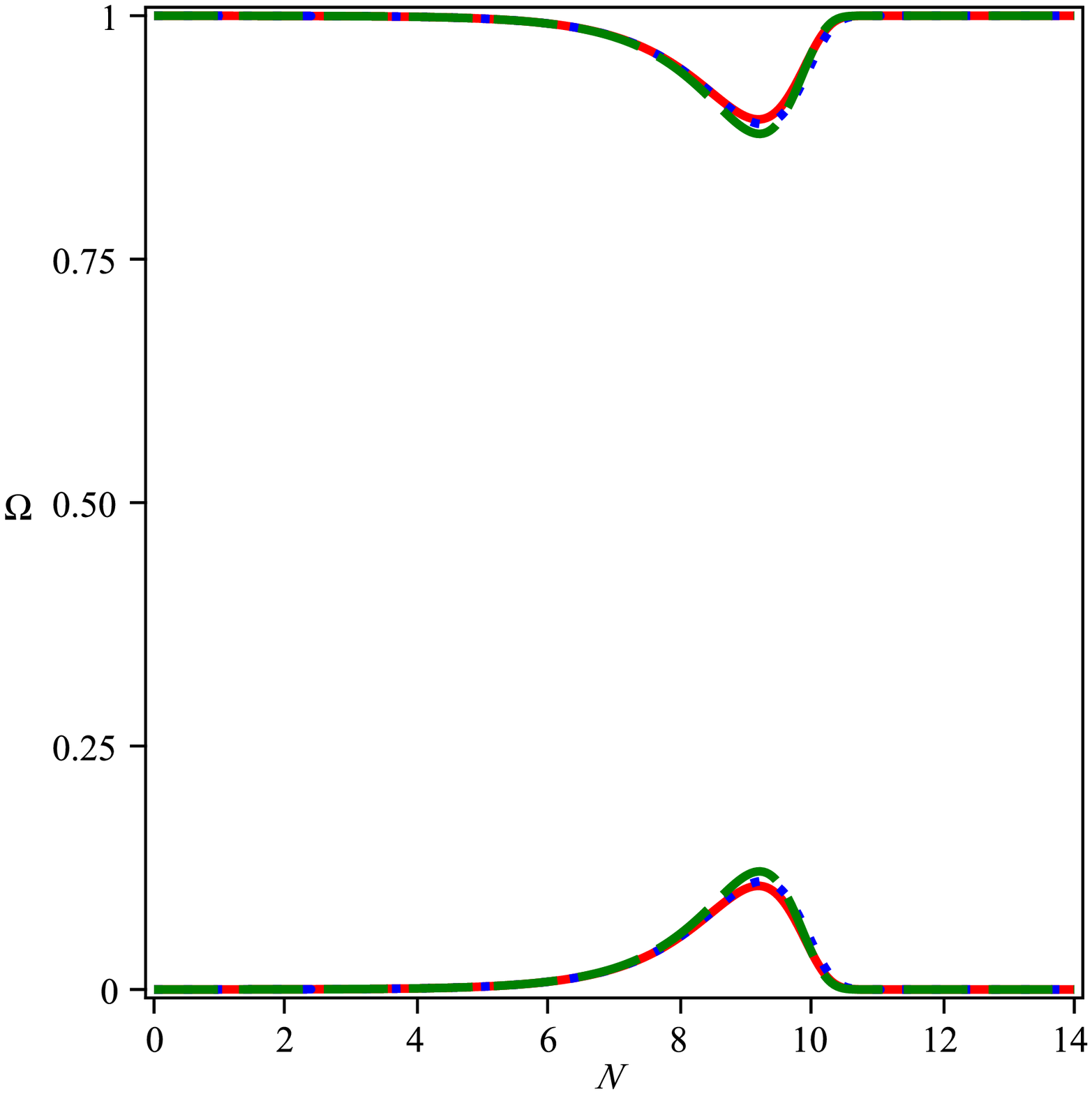}
\hspace{1cm}
\includegraphics[width=0.4\linewidth]{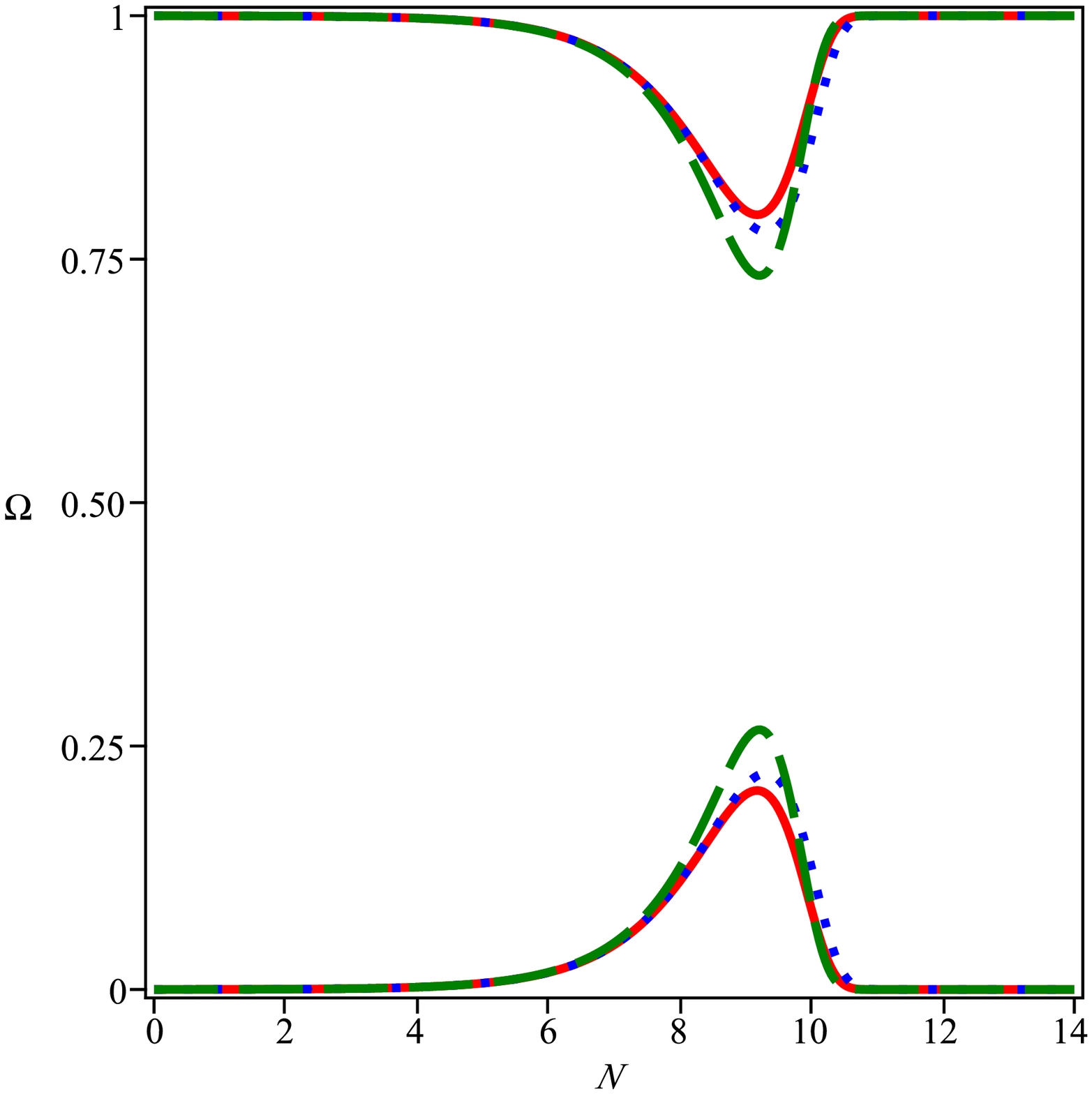}
\caption{ The evolution of $\Omega_{\rm r}$ and $\Omega_\sigma$ for $\Gamma/H_0=10^{-8}$.
The left panel is for $f_{\rm d}$, the weighted ratio of the curvaton energy density to the total energy density at the time of curvaton decay, equal to $0.08$ and the right panel for $f_{\rm d} = 0.16$. In each plot, the upper curves are $\Omega_{\rm r}$ while the lower curves are  $\Omega_\sigma$. The solid red curves are the full numerical solution, the blue dotted curves the analytic solutions, Eqs. (\ref{sol-sigma}), and the green dashed curves the analytic solutions  in the $f_{\rm d} \ll 1$ limit, Eqs.~(\ref{limit}).  As expected, for small values of $f_{\rm d}$ the agreement between the numerical and the analytical solutions is very good.  }
\label{fig1}
\end{figure}

In Fig.~\ref{fig1} we have plotted the full numerical solutions for $\Omega_\sigma(N)$ and  $\Omega_{\rm r} (N)$ and compared them with our analytical solutions Eqs.~(\ref{sol-sigma}) and (\ref{limit}).  As can be seen, for small values of $\Omega_\sigma|_{\mathrm{dec}}$, or equivalently small values of  $f_{\rm d}$, the agreement between our analytical solutions and the full numerical solutions is very good.

Finally we obtain an estimate for $N_{\rm d}$, the time of curvaton decay.  A good criteria to define $N_{\rm d}$ in the sudden decay limit is when
the last term in  Eq.~(\ref{X-eq}) becomes comparable to the second term and $X'(N_{\rm d})=0$, and 
the ratio $\Omega_\sigma/\Omega_{\rm r}$ reaches its maximum.
This gives $\Gamma (1+ X (N_{\rm d})) = H(N_{\rm d})$. In the limit where $X(N_{\rm d})\ll 1$, or equivalently $f_{\rm d} \ll 1$, this reduces to the standard result that $H(N_{\rm d}) \simeq \Gamma$ or  
\ba
\label{Nd-eq}
N_{\rm d} \simeq \frac{1}{2} \ln \left( \frac{H_0}{\Gamma } \right) \, ,
\quad \quad \quad \quad       \left( f_{\rm d}\ll 1 \right) \,.
\ea 
Inserting this expression for $N_{\rm d}$ into the definition of $f_{\rm d}$ in Eq. (\ref{fd-def}) yields
\ba
\label{fd-eq}
f_{\rm d} \sim \frac{3 p}{4} \left(\frac{H_0}{\Gamma} \right)^{1/2}  \,,
\quad \quad  \left( f_{\rm d}\ll 1 \right) \,.
\ea
In general, when $f_{\rm d} \gtrsim 1/2$ then Eq.~(\ref{Nd-eq}) receives corrections and one has to find $N_{\rm d}$ by solving $X'=0$ with $X$ given by Eq.~(\ref{X-sol2}).

\section{Perturbations}
\label{pert_sect}

In this section we study the perturbed Einstein and fluid equations.
The perturbed metric line element is 
\cite{Mukhanov:1990me,Malik:2004tf,Malik:2008im}
\ba
\label{metric-pert}
{\rm d} s^2 = -(1 + 2 \phi) {\rm d} t^2 + 2 a B_{, i}{\rm d} t {\rm d} x^i + a^2 \left[ (1- 2 \psi) \delta_{ij} + 2 E_{, ij}
\right]  {\rm d} x^i {\rm d} x^j \,.
\ea
The perturbed Einstein equations are then
\ba
\label{Ein1}
\Psi' + \frac{5+3 \omega}{2} \Psi + \frac{k^2}{3 a^2 H^2} \Psi + \frac{3}{2} (1+\omega) \zeta=0 \, ,
\ea
and
\ba
\label{Ein2}
\Psi' + \frac{5+3 \omega}{2} \Psi  - \frac{3}{2} (1+ \omega) {\cal R} =0 \,,
\ea
where the time-dependent equation of state $w$ is given by
\ba
\label{omega}
{ w} \equiv \frac{P}{\rho} = \sum_\alpha \frac{ \omega_\alpha\, \rho_\alpha}{\rho} \,,
\ea
in which $\omega_\alpha$ is the equation of state for each fluid given
by $\omega_\alpha = p_\alpha/\rho_\alpha$ with $\omega_\sigma =0$ and
$\omega_r = 1/3$.

Here we have defined the Bardeen potential, or curvature perturbation
on uniform shear hypersurfaces, as
\ba
\Psi\equiv \psi - H(B-\dot E)\,,
\ea
and the curvature perturbations on uniform density slices
$\zeta$, and on comoving hypersurfaces, ${\cal R}$, respectively as 
\ba
\label{zeta}
\zeta \equiv -\psi - \frac{ H\delta \rho}{\dot \rho}  \, ,\qquad
\label{R}
{\cal R} \equiv \psi - HV \,,
\ea
where $V \equiv a (v + B)$,
and $v$ is the total scalar velocity potential.

Equations~(\ref{Ein1}) and (\ref{Ein2}) can be combined to give
\ba
\label{Ein3}
\frac{k^2}{3 a^2 H^2}\Psi = -\frac{3}{2} (1+ \omega) (\zeta + {\cal R} ) \, .
\ea
In particular, we see that on large scale where $k/aH \rightarrow 0$,
$\zeta \simeq -{\cal R}$.

The equations of motion for each fluid are
\ba
\label{delta-rho-eq3}
\delta  \rho_\alpha' + 3 (\delta \rho_\alpha + \delta p_\alpha) - 3 (\rho_\alpha + p_\alpha)  \psi' - \frac{k^{2}}{a^{2}H}(\rho_\alpha + p_\alpha)  (V_\alpha+\sigma_s ) - \frac{1}{H}(Q_\alpha \phi + \delta Q_\alpha)
&=&0 \,,\\
 \label{V-eq3}
 V'_\alpha  + \left[ \frac{Q_\alpha}{H(\rho_\alpha + p_\alpha)} (1+ c_\alpha^2) - 3  c_\alpha^2  \right] V_\alpha +\frac{ \phi }{H}+ \frac{1}{H(\rho_\alpha + p_\alpha)}\left[ \delta p_\alpha - Q_\alpha V
 \right]  &=&0 \,,
 \ea
where the sound speed for each fluid is defined by $c_\alpha^2 = \dot p_\alpha/\dot \rho_\alpha$.  We shall assume further below that
 each fluid is intrinsically adiabatic so $p_\alpha = p_\alpha (\rho_\alpha)$ and $\delta p_\alpha = c_\alpha^2 \delta \rho_\alpha $. Also $v_\alpha$ is the scalar velocity potential for each fluid
 and  $V_\alpha \equiv a (v_\alpha +B)$.

In order to express the fluid equations, Eqs.~(\ref{delta-rho-eq3}) and (\ref{V-eq3}), in gauge invariant form, we define the curvature perturbations
$\zeta_\alpha$ and ${\cal R}_\alpha$ for each fluid as
\ba
\label{zeta-alpha}
\zeta_\alpha \equiv -\psi - H \frac{\delta \rho_\alpha}{\dot \rho_\alpha} \,,
\ea
and
\ba
\label{R-alpha}
{\cal R}_\alpha \equiv \psi  - H V_\alpha  \, .
\ea
 One can cast the perturbed fluid equations, Eqs.~(\ref{delta-rho-eq3}) and (\ref{V-eq3}), into \cite{Malik:2004tf}
\ba
\label{zeta-alpha-eq}
\zeta_\alpha' = -\frac{H' Q_\alpha}{H^2 \rho_\alpha'}
\left( \frac{\delta \rho_\alpha}{\rho_\alpha'} - \frac{\delta \rho}{\rho'} \right)
- \frac{1}{H \rho_\alpha'} \left( \delta Q_\alpha - \frac{Q_\alpha'}{\rho_\alpha'} \delta \rho_\alpha \right) + \frac{k^2}{3 a^2 H^2} \left[ \Psi - \left( 1- \frac{Q_\alpha}{H \rho_\alpha'}
 \right) {\cal R_\alpha}  \right] \,,
\ea
and
\ba
\label{R-alpha-eq}
{\cal R}_\alpha' = \left(  \frac{Q_\alpha}{H( \rho_\alpha + p_\alpha)}  -  \frac{H'}{H}\right)
 \left( {\cal R} -{ \cal R}_\alpha  \right) - \frac{\rho_\alpha'}{\rho_\alpha + p_\alpha} c_\alpha^2 \left({ \cal R}_\alpha + \zeta_\alpha
\right) \, .
\ea
In the absence of energy transfer between the fluids, $Q_\alpha = \delta Q_\alpha=0$, we see that $\zeta_\alpha$ for each fluid is constant on super-horizon scales
\cite{Wands:2000dp}. Note that in deriving Eqs.~(\ref{zeta-alpha-eq}) and (\ref{R-alpha-eq})
we have assumed that each fluid is intrinsically adiabatic so $p_\alpha = p_\alpha (\rho_\alpha)$
and $\delta p_\alpha = c_\alpha^2 \delta \rho_\alpha $.

We also note that $\zeta$ and ${\cal R}$ defined in Eq.~(\ref{zeta}) 
can be written as the weighted sum of $\zeta_\alpha$ and ${\cal R}_\alpha$,
\ba
\label{zeta-2}
\zeta = \sum_\alpha \frac{\dot \rho_\alpha}{\dot \rho} \zeta_\alpha  
= \sum_\alpha \frac{(1+ \omega_\alpha) \rho_\alpha}{(1+ \omega) \rho  }
{\zeta}_\alpha \,,
\ea
and
\ba
\label{R-2}
{\cal R} = \sum_\alpha \frac{(1+ \omega_\alpha) \rho_\alpha}{(1+ \omega) \rho  }
{\cal R}_\alpha \, .
\ea

So far we have not specified the perturbations in the energy transfer
$\delta Q_\alpha$. Following Ref.~\cite{wands1}, we assume that the
decay rate $\Gamma$ is fixed by the microphysics so $\delta \Gamma=0$
and therefore
\ba
\delta Q_\sigma = - \Gamma \delta \rho_\sigma \,, \quad
 \delta Q_\gamma =  \Gamma \delta \rho_\sigma \, .
\ea

The system of equations in terms of $\{ \zeta_\sigma, {\cal R}_\sigma,
\zeta, {\cal R} \}$ is given by

\ba
\label{zeta-cur1}
\zeta_\sigma' &=& \frac{(3+ \Omega_\gamma) \Gamma }{2 (\Gamma + 3 H) }
(\zeta - \zeta_\sigma)  + \frac{k^2}{3 a^2 H^2}  \Psi - \frac{k^2}{ a^2 H^2}\left(\frac{H}{3H+\Gamma}\right){\cal R}_\sigma  \,, \\
\label{zeta1}
\zeta' &=& \frac{(\Gamma + 3 H) \Omega_{\sigma} }{ H(3+ \Omega_\gamma) }
(\zeta_\sigma - \zeta)
  + \frac{k^2}{3 a^2 H^2} (\Psi - {\cal R} )  \,, \\
\label{R-cur1}
{\cal R}_\sigma' &=& - \left( \frac{ \Gamma}{H} + \frac{H'}{H}
\right) \left({\cal R} - \frac{3 (1+ \Omega_\gamma)}{3 + \Omega_\sigma} {\cal R}_\sigma\right)  \,, \\
\label{R1}
{\cal R}' &=&  \left( \frac{H'}{H} + \frac{4 H \Omega_\gamma - \Gamma \Omega_\sigma }{H (3 +  \Omega_\gamma)} \right) {\cal R}
+ \left( 1+ \frac{H'}{H} \right) \zeta - \frac{(\Gamma + 3 H) \Omega_{\sigma}}{H(3+ \Omega_\gamma)} \zeta_\sigma - \frac{k^2}{3 a^2 H^2}  \Psi \,.
\ea
Note that this is a closed system of equation for $\{ \zeta_\sigma, {\cal R}_\sigma,
\zeta, {\cal R} \}$ and that $\Psi$ can be eliminated from these equations
using Eq.~(\ref{Ein3}) in terms of $\zeta $ and ${\cal R}$.

Alternatively, one may write the system of equations in terms of $\{ \zeta_{\rm r}, {\cal R}_{\rm r}, \zeta, {\cal R} \}$ or $\{ \zeta_{\rm r}, {\cal R}_{\rm r},
\zeta_\sigma, {\cal R}_\sigma \}$ as given in Appendix \ref{fluid-eqs}.

\section{Analytic Calculation of Bardeen potential}
\label{bardeen_sect}

In this section we provide the analytical solutions for the Bardeen
potential $\Psi$ in different limits. As can be seen, the system of
Eqs.~(\ref{zeta-cur1})-(\ref{R1}) is too complicated to be handled
analytically for all modes. However analytical solutions for $\Psi$
can be obtained in some limiting situations. In the next two
subsections we consider modes which are super-horizon at the time of
curvaton decay, $k< a(N_d) H(N_d)$, and modes which are always
sub-horizon during curvaton evolution corresponding to $k > H_0$ where
$H_0$ is the Hubble constant at the end of reheating when $a(N=0)=1$.

\subsection{Super-horizon modes}
\label{supe-sub}

Here we provide the solution for the modes which are super-horizon at
the time of curvaton decay and re-enter the horizon during the second
radiation stage. First we consider the epoch before the curvaton
decays, $N< N_{\rm d}$. In the sudden decay limit, for the
super-horizon modes from Eqs.~(\ref{zeta-cur}) and (\ref{zeta-gamma})
it can be shown that $\zeta_{\rm r}'\simeq 0$ and $\zeta_\sigma'\simeq
0$. As expected, on super-horizon scales both $\zeta_\sigma$ and
$\zeta_{\rm r}$ remain frozen so one can approximate $\zeta_\sigma$
with its value at the time of horizon crossing during inflation
\ba
\zeta_\sigma(N) \simeq \zeta_{\sigma_*}   \, .
\ea
Also to further simplify
the analysis, we consider the conventional curvaton mechanism where
$\zeta_{\gamma \, {\rm in}}=0$ and there is no initial radiation
perturbation, corresponding to entropic initial conditions.  In this
limit, either by solving Eq.~(\ref{zeta2}) or using the definition
(\ref{zeta-2}), we have 
\ba
\label{zeta-R-sigma}
\zeta = \frac{3 \Omega_\sigma}{4- \Omega_\sigma} \zeta_{\sigma_*} \,.
\ea
Substituting this into Eq.~(\ref{Ein1}) and noting that $w =
\Omega_{\rm r}(N)/3$ and using Eq.~(\ref{solbackH}), results in 
\ba 
\label{Psi-eq0}
\Psi' + \frac{1}{2} \left( 5 + \frac{1}{1+ p\, e^N } \right) \Psi+ \frac{3 p\,
e^{N} \zeta_{\sigma_*}}{2(1+ p\, e^N)} =0  \,.
\ea 
This can easily be solved with the result 
\ba
\label{Psi-anal}
\Psi(N) =\left[C\,  \sqrt{1+ p\, e^N}  - \frac{3 \zeta_{\sigma_*}}{5 p^3} \left( 16 + 8 p\, e^N - 2 p^2 \, e^{2N} + p^3 e^{3N} \right) \right] e^{-3 N} \,.
\ea
Here $C$ is a constant of integration which is obtained by 
matching this solution to the value
of $\Psi$ at the end of inflation, which gives 
\ba
\label{C-def}
C = \frac{1}{\sqrt{1+p} } \left[ \Psi(0) +  \frac{3 \zeta_{\sigma_*}}{5 p^3} \left( 16 + 8 p - 2 p^2  + p^3  \right) \right] \,  
\simeq \frac{48 \zeta_{\sigma_*}}{5 p^3} \, .
\ea
To obtain the second approximate relation we used $p \ll 1$ and $\Psi(0) \simeq 0$ which is a good approximation for all modes at the end of inflation \cite{Lyth:2005ze}.

Having obtained $\Psi$ during curvaton evolution, we now find $\Psi$ after the curvaton decays. The governing equation for $\Psi$  during the radiation domination stage with $\omega_{\rm r}=1/3$ has the standard form 
\ba
\label{Psi-rad-eq}
\Psi'' + 3 \Psi' + \frac{k^2}{3 a^2 H^2} \Psi =0  \, .
\ea
Using
\ba
\label{H-rad}
H = H_{{\rm d}} e^{-2 (N- N_{{\rm d}})} \,,
\ea
during the radiation era  this leads to a solution in terms of Bessel functions,
\ba
\label{Psi-second}
\Psi(N) = e^{-3N/2} \left[ c_1 J_{3/2} \left( \bar k e^{N} \right) +  c_2 Y_{3/2} \left( \bar k e^{N}\right) \right] \,,
\ea
in which $c_1$ and $c_2$ are two constants of integration and 
\ba
\label{kbar}
\bar k \equiv \frac{ k e^{-2 N_{{\rm d}}}}{\sqrt 3 H_{\rm d}} \, .
\ea
In the limit where the curvaton makes a sub-dominant contribution to the total energy density at the time of decay, corresponding to $f_{\rm d} \ll 1 $, one can use Eqs.~(\ref{solbackH}) and (\ref{Nd-eq}) to obtain 
$H_{\rm d} \simeq \Gamma$ and $\bar k \sim k/H_0$. Also the condition for the mode $k$ to be superhorizon at the time of curvaton decay, $k> H_{\rm d} e^{N_{\rm d}}$, translates into 
$k > \sqrt{\Gamma H_0}$.

We can now determine the constants of integration $c_1$ and
$c_2$, by requiring that both $\Psi$ and $\Psi'$
are continuous at the time of curvaton decay, $N=N_d$. This gives
\ba
 \label{c-12}
 c_1 &=& \frac{\pi}{2}  \left[ \left( \bar k e^{5 N_{\rm d}/2} Y'_{3/2}(\bar k e^{N_{\rm d}}) - \frac{3}{2} e^{3N_{\rm d}/2} Y_{3/2}(\bar k e^{N_{\rm d}})  \right) \Psi(N_{\rm d}) -  e^{3N_{\rm d}/2} Y_{3/2}(\bar k e^{N_{\rm d}}) \Psi'(N_{\rm d})  \right] \,,\\
  c_2 &=& \frac{\pi}{2}  \left[ \left( -\bar k e^{5 N_{\rm d}/2} J'_{3/2}(\bar k e^{N_{\rm d}}) + \frac{3}{2} e^{3N_{\rm d}/2} J_{3/2}(\bar k e^{N_{\rm d}})  \right) \Psi(N_{\rm d}) +  e^{3N_{\rm d}/2} J_{3/2}(\bar k e^{N_{\rm d}}) \Psi'(N_{\rm d})  \right]\,.
 \ea
Note that in the above expressions $\Psi(N_{\rm d})$ and $\Psi'(N_{\rm d})$ are calculated from the solution
obtained from the period before the curvaton decay, Eq.~(\ref{Psi-anal}).

By construction, we know that $a(N_{\rm d}) H(N_{\rm d}) < k$,  so $\bar k e^{N_{\rm d}} \sim k e^{-N_{\rm d}}/H_{\rm d} <1$. Using the small argument limits of the Bessel functions we find  that $c_2 \ll c_1$ and therefore
\ba
\label{psin}
\Psi(N) \simeq c_1 e^{-3N/2} J_{3/2} (\bar k e^N) \,,  \quad    (N \geq N_d) \, .
\ea

\begin{figure}
\includegraphics[width=0.4\linewidth]{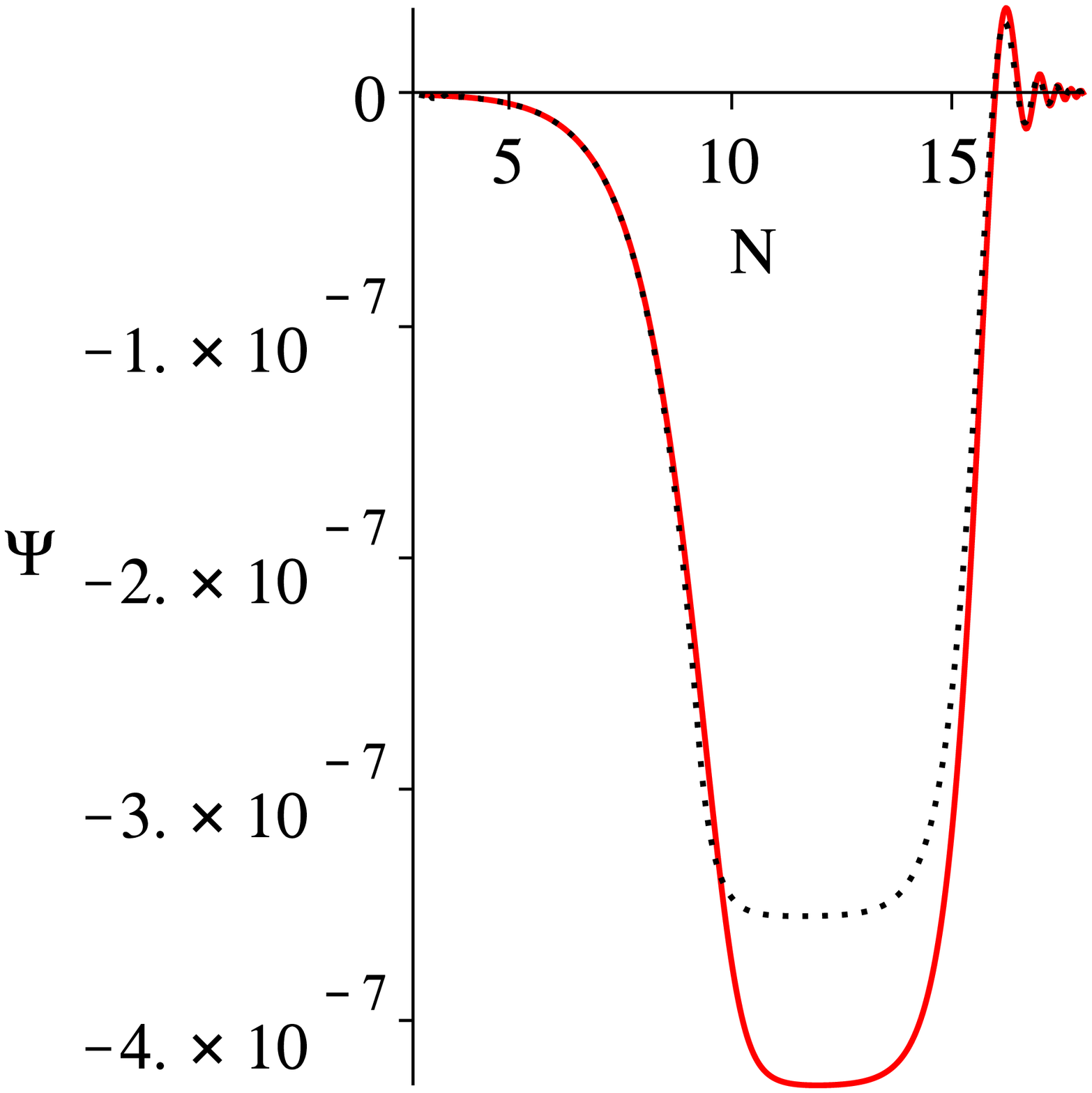}
\hspace{1cm}
\includegraphics[width=0.4\linewidth]{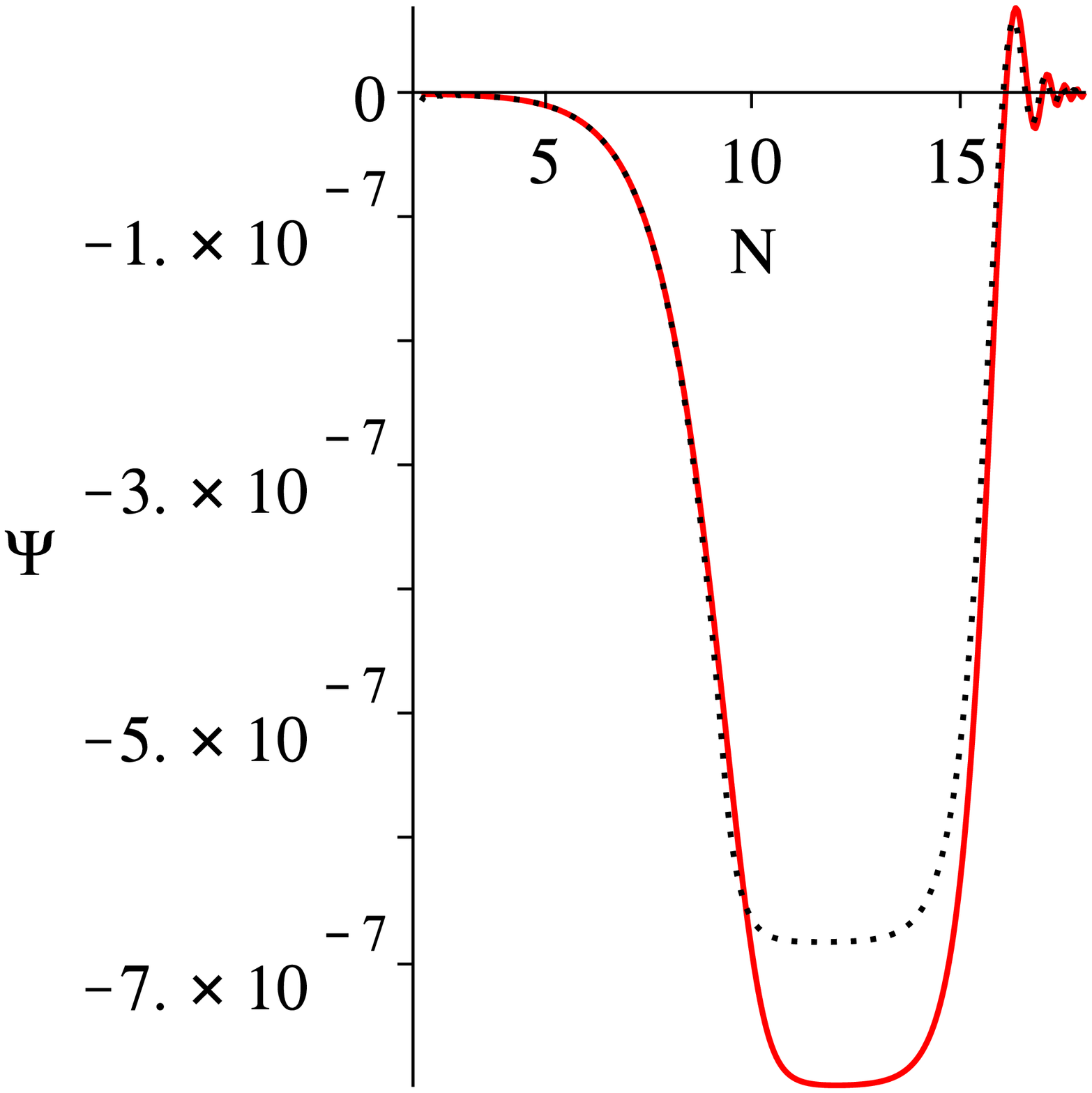}
\caption{The evolution of the Bardeen potential $\Psi$ for super-horizon modes with $f_{\rm d} = 0.08$ and 
$0.16$ (left and right plots respectively). In both plots $\Gamma/H_0 =10^{-8}$ and the red curve is the full numerical result while
the black dotted curve is our analytical solution, Eq.~(\ref{Psi-second}).   In both plots, the existence of the plateau, Eq.~(\ref{plateau}), for the modes in the range $N_{\rm d} < N <N_k$ is evident. As expected from Eq.~(\ref{c1-limit1}), the larger $f_{\rm d}$ is, the larger the amplitude of $\Psi$ is.
}
\label{super-fig}
\end{figure}

Now suppose these modes (with $\bar k e^{N_{\rm d}} <1$) re-enter the horizon during the second radiation era at the time $N= N_k$.  The value of $N_k$ can be estimated by calculating when the argument of the Bessel function in Eq.~(\ref{psin}) becomes of order unity and $J_{3/2} (\bar k e^N)$
starts to oscillate. This yields $\bar k e^{N_k}=1$ or $N_k = -\ln \bar k$. During the period 
$N_d < N< N_k$ in which the mode is still super-horizon during the second radiation stage 
we can use the small argument limit of the Bessel function $J_\nu (x) \simeq (x/2)^\nu/\Gamma(\nu+1)$ (here $\Gamma(\nu+1)$ is the gamma-function)
to obtain
\ba
\label{plateau}
\Psi(N)  \simeq  \frac{c_1}{\Gamma(5/2)} \left(\frac{\bar k}{2} \right)^{3/2} \,, \quad
 N_d \leq N < N_{k} \, .
\ea
This is a very interesting result; for $N_{\rm d}
< N < N_{k}$, $\Psi$ is constant with the value given by
Eq.~(\ref{plateau}). Numerical evolution of $\Psi$, shown in Fig.~\ref{super-fig},  verifies the existence of
this plateau.\\

During the epoch $N>N_k$, after the mode $k$ has re-entered the horizon, 
the Bessel function in Eq.~(\ref{psin}) oscillates rapidly. 
Using the large argument limit of the Bessel function we obtain
\ba
\Psi(N) \simeq \sqrt{\frac{2}{\pi \bar k}} e^{-2 N} c_1\,  \cos \left( \bar k e^N + \frac{\pi}{4} \right) \,  
\,,  \quad    (N > N_k) \, .
\ea

As we shall see in the following section in order to study whether PBHs are overproduced, we need to estimate $\Psi(N_{\rm d})$ and $c_1$.  As mentioned before, the key parameter in controlling the amplitude of $\Psi(N)$ during the second radiation era is $f_{\rm d}$.
Here we calculate $c_1$ for two extreme cases  (i): $f_{\rm d} \ll 1$ corresponding to $p e^{N_{\rm d}} \ll 1$ and (ii): $f_{\rm d} \gtrsim 1/2$ corresponding to $p e^{N_{\rm d}} \ll 1$.

For the limit $f_{\rm d} \ll1$, from Eq.~(\ref{Psi-anal}) we obtain
\ba
\Psi(N_{\rm d}) \simeq -\frac{3}{8} p e^{N_{\rm d}} \zeta_{\sigma_*} \simeq -\frac{f_{\rm d}}{2} \zeta_{\sigma_*} \,, \quad (f_{\rm d} \ll 1) \,,
\ea
where to obtain the second approximation, Eqs.~(\ref{Nd-eq}) and  (\ref{fd-eq}) have been used. Similarly, using Eq.~(\ref{Psi-eq0}) to eliminate $\Psi'$ we obtain
\ba
\Psi'(N_{\rm d}) \simeq -\frac{f_{\rm d}}{2} \zeta_{\sigma_*} \simeq \Psi(N_{\rm d})  
\,, \quad (f_{\rm d} \ll 1) \,.
\ea
As a result
\ba
\label{c1-limit1}
c_1 \simeq \frac{\pi}{2} {\bar k}^{-3/2} \left(\,  3 \Psi(N_{\rm d}) + \Psi'(N_{\rm d}) \,   \right)
\simeq -\sqrt{2 \pi} f_{\rm d} {\bar k}^{-3/2} \zeta_{\sigma_*}  \,, \quad (f_{\rm d} \ll 1) \,.
\ea
We have checked this numerically and our analytical estimation of $c_1$ is in good agreement with its numerical value.

On the other hand, in the limit where $p e^{N_{\rm d}} \gg 1$, so $f_d \gtrsim 1/2$,
one finds
\ba
\label{Psi-Nd}
\Psi(N_d) \simeq -\frac{3}{5} \zeta_\sigma \quad , \quad
\Psi'(N_d) \simeq -\frac{5}{2} \Psi(N_d) - \frac{3}{2} \zeta_\sigma \simeq 0 \,.
\ea
%
Inserting these results into the expression for $c_1$ in Eq.~(\ref{c-12}) yields
\ba
\label{c1-limit2}
c_1 \simeq - \frac{9}{5\, { \bar k}^{3/2}} \sqrt{\frac{\pi}{2}} \zeta_{\sigma_*} 
\,, \quad \left(f_{\rm d} \gtrsim \frac{1}{2} \right) \,.
\ea
This indicates that for large enough $f_{\rm d}$, the amplitude of $\Psi$ during the second radiation dominated epoch is nearly independent of $f_{\rm d}$.

In Fig.~\ref{super-fig} we have plotted $\Psi(N)$ for the super-horizon modes with $f_{\rm d} \ll1$ and $f_{\rm d} \gtrsim 1/2$. The main result here is that the amplitude of $\Psi$ increases during the final radiation dominated era due to curvaton dynamics. However, the increase in
amplitude of $\Psi$ is less efficient for small $f_{\rm d}$.
We shall see whether this will have implications for primordial black hole formation in Section \ref{pbh_sect} below.

\subsection{Sub-horizon modes}
\label{sub-section}

In this sub-section we calculate $\Psi$ for the sub-horizon modes. In general, it is not easy to solve the system of equation in this limit. A particular limit which may be handled semi-analytically is the case where the term $(k/a H)^2 \gg 1$ throughout curvaton dynamics, corresponding to $k> H_0$. These are modes which are sub-horizon during the entire curvaton dynamics.

Differentiating Eq.~(\ref{Ein1}) and using Eq.~(\ref{zeta1}) and Eq.~(\ref{Ein2}) to 
to eliminate $\zeta'$ and ${\cal R}$ we obtain the following second order 
differential equation for $\Psi$
\ba
\label{Psi-sec}
\Psi'' + \frac{\Omega_\sigma^2 - 8 \Omega_\sigma + 24}{2 (4 -\Omega_\sigma)} \Psi'
+ \frac{4 (1 -\Omega_\sigma)}{3 (4- \Omega_\sigma)} \frac{k^2}{a^2 H^2} \Psi=0 \,.
\ea
In the limit $\Omega_\sigma=0$, this reduces to the standard equation for
$\Psi$ in a radiation dominated background given in Eq.~(\ref{Psi-rad-eq}).

\begin{figure}
\includegraphics[width=0.45\linewidth]{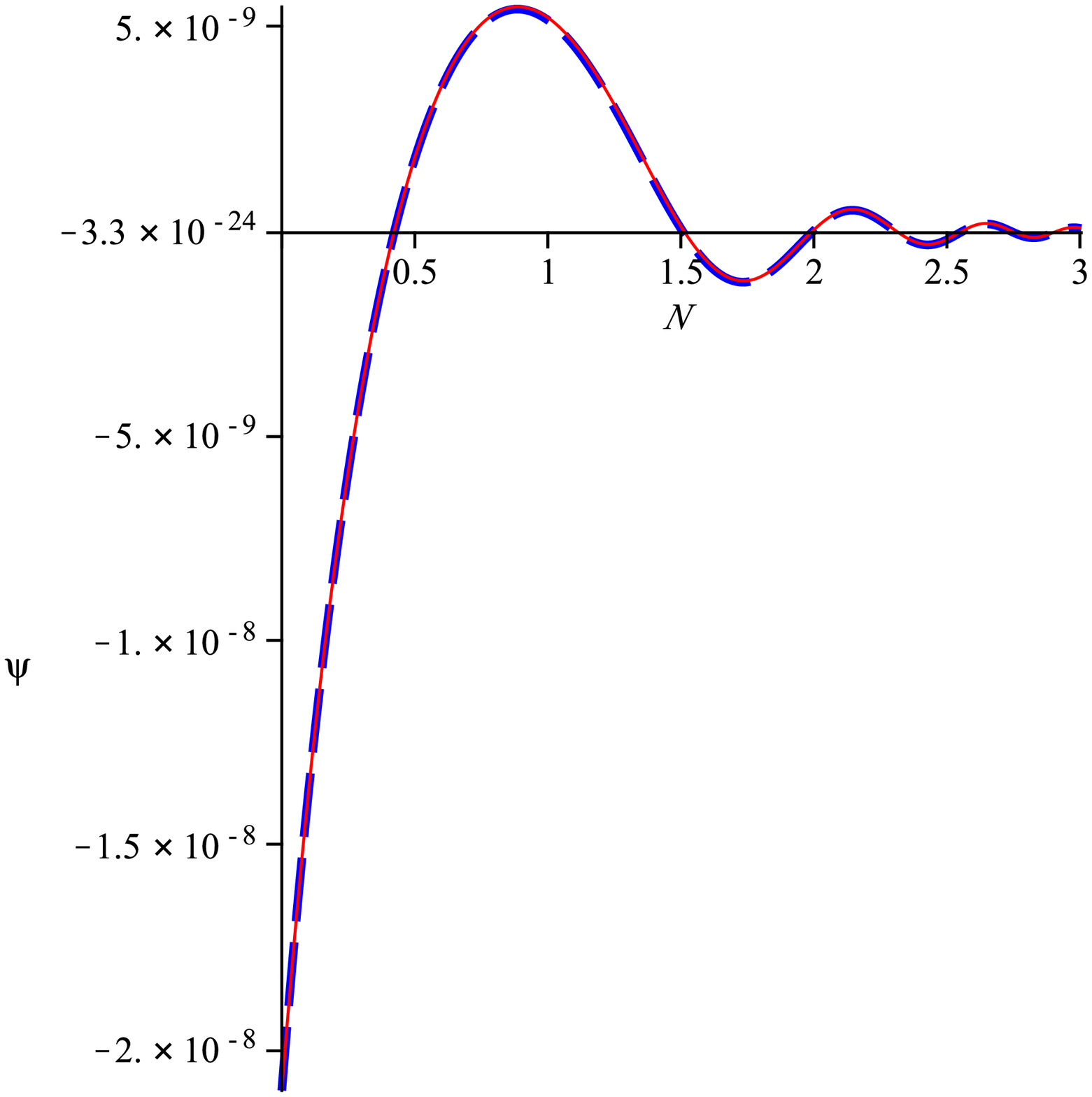}
\hspace{0.5cm}
\includegraphics[width=0.45\linewidth]{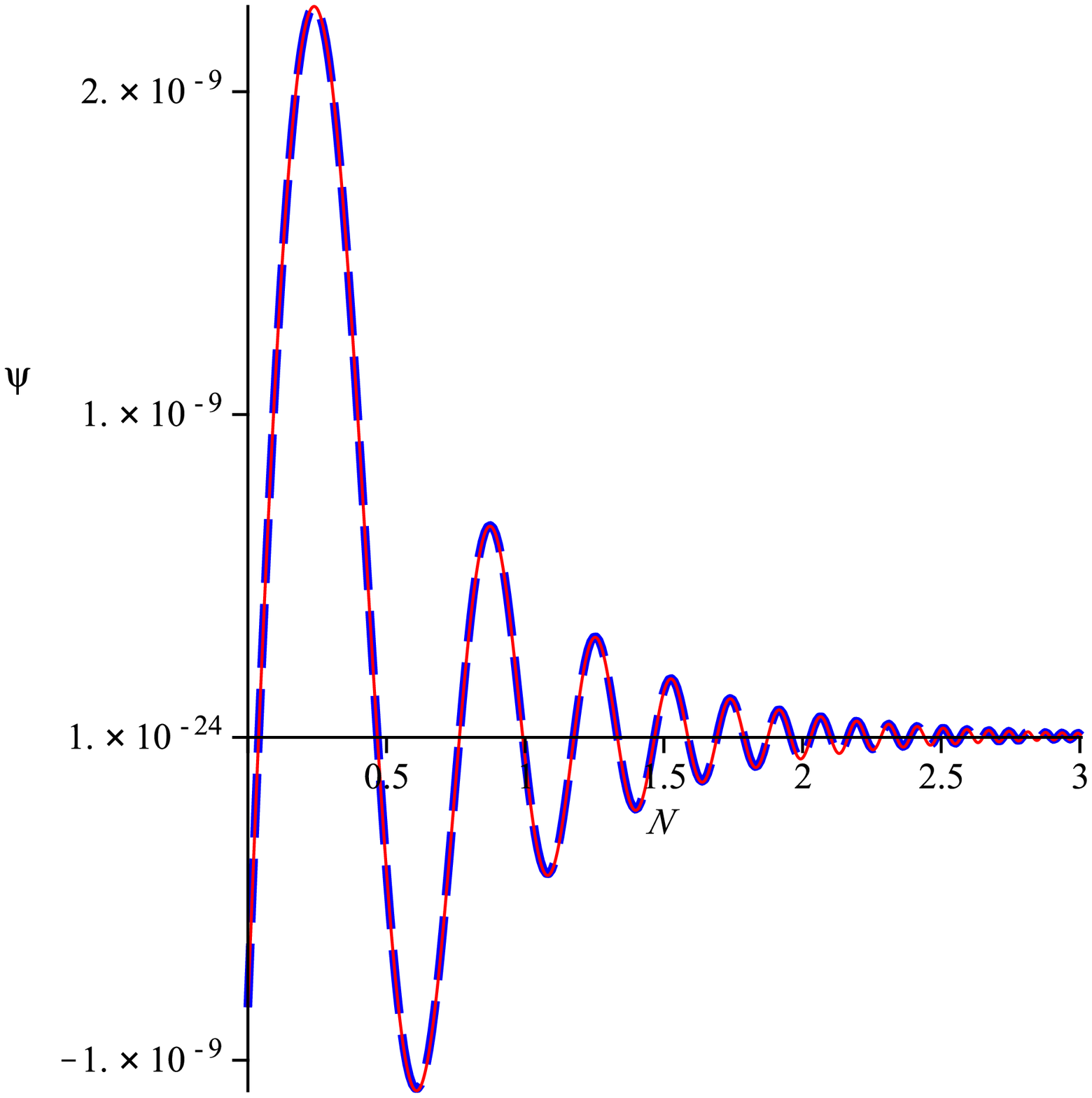}
\caption{The evolution of the Bardeen potential $\Psi$ for sub-horizon modes with $k/ H(0)=2$ and $10$ in the left and right panels respectively, for $f_{\rm d} = 0.16$ and $\Gamma/H_0 =10^{-8}$.
 In both plots, the thick dashed blue curve
is the full numerical result while the thin solid red curve is the numerical solution obtained from
Eq.~(\ref{Psi-sec}). 
}
\label{sub-fig}
\end{figure}

Equation (\ref{Psi-sec}) can not be solved analytically, as far as we know. In Fig.~\ref{sub-fig} we have plotted $\Psi$ for different values of $k$ corresponding to modes which are deep inside the horizon at the end of inflation.  In principle one could solve Eq.~(\ref{Psi-sec}) semi-analytically and find the values of $\Psi(N_{\rm d})$ and $\Psi'(N_{\rm d})$ and insert them into Eq.~(\ref{psin}) to find $\Psi$ in the final radiation dominated era.
With an analytical solution for $\Psi$ in this regime one could then calculate the abundance of PBHs formed from sub-horizon fluctuations, c.f.~Refs.~\cite{Lyth:2005ze, Zaballa:2006kh}. This is beyond the scope of this work however, and we focus in the next section on the `standard' case of PBHs forming when super-horizon modes reenter the horizon, using the analytic solution for $\Psi$ from subsection ~\ref{supe-sub} in the final radiation dominated era, i.e. Eq.~(\ref{psin}),
with $c_1$ given by Eqs.~(\ref{c1-limit1}) and (\ref{c1-limit2}).

\section{ Primordial Black Hole Formation}
\label{pbh_sect}


Primordial black holes (PBHs) are a powerful tool for constraining models of the early Universe.
Due to their gravitational effects
and the consequences of their evaporation there are tight constraints on the number of PBHs that form (see e.g. Refs.~\cite{Josan:2009qn,Carr:2009jm}).  These abundance constraints can be used to constrain the primordial power spectrum,
and hence models of the early Universe, on scales far smaller than those
probed by cosmological observations. We found in Sec.~\ref{bardeen_sect} that the power spectrum is enhanced during the radiation dominated period after curvaton decay. In this section we therefore investigate whether this enhancement is sufficiently large to lead to PBH over production.\\

A region will collapse to form a PBH, with mass of order the horizon mass at that epoch, if the smoothed density contrast,
in the comoving gauge, at horizon crossing ($R=c_{\rm s} (a
H)^{-1}$ where $c_{\rm s}=1/\sqrt{3}$ is the sound speed), $\delta_{\rm com}(R= 1/a H) \equiv \delta_{\rm hor}(R)$, satisfies the
condition $\delta_{\rm hor}(R) \geq \delta_{\rm c}$~\cite{Carr:1985wss},
where $\delta_{\rm c} \sim 1/3$~\footnote{It was previously thought that there was an upper limit on the size of fluctuations which form PBHs, with larger fluctuations forming a separate closed universe, however Kopp et al.~\cite{Kopp:2010sh} have recently shown that this is in fact not the case.}. 
The fraction of the energy density of the Universe contained in
regions dense enough to form PBHs is then given, as in
Press-Schechter theory~\cite{Press:1973iz}, by
\begin{equation}
\beta= 2 \frac{M_{\rm PBH}}{M_{H}}
\int_{\delta_{\rm c}}^{\infty} 
P(\delta_{\rm hor}(R)) \,{\rm d}\delta_{\rm hor}(R) \,.
\label{presssch}
\end{equation}
Assuming that the initial primordial perturbations are Gaussian, the
probability distribution of the smoothed density contrast,
$P(\delta_{\rm hor}(R))$, is given by (e.g.~Ref.~\cite{LL})
\begin{equation}
\label{prob}
P(\delta_{\rm hor}(R))=\frac{1}{\sqrt{2\pi} \sigma_{\rm hor}(R)} \exp{ \left(
    - \frac{\delta_{\rm hor}^2(R)}{2 \sigma_{\rm hor}^2(R)} \right)} \,,
\end{equation}
where $\sigma(R)$ is the mass variance
\begin{equation}
\label{variance1}
\sigma^2(R)=\int_{0}^{\infty} W^2(kR)\mathcal{P}_{\delta}(k, t)\frac{\,{\rm d}k}{k}.
\end{equation}
Here $W(kR)$ is the Fourier transform of
the window function used to smooth the density contrast, which we take to be Gaussian, so that $W(kR)= \exp{(-k^2 R^2/2)}$ and $\mathcal{P}_{\delta}(k, t)$ is the power spectrum of the comoving density contrast,
\begin{equation}
\label{powerlawps}
{\cal{P}}_{\delta} (k,t) \equiv \frac{k^3}{2 \pi^2} \langle
        |\delta_{\rm com}(k,t)|^2 \rangle \,.
\end{equation}
Inserting the expression for the probability distribution, \eq{prob}, into the Press-Schechter expression for the initial PBH abundance, \eq{presssch}, gives
\begin{equation}
\label{densitypara}	
\beta =  \frac{2 }{\sqrt{2\pi}\sigma_{\rm hor}(R)}
\int_{\delta_{\rm c}}^{\infty} \exp{\left(- \frac{\delta^2_{\rm hor}(R)}
    {2 \sigma_{\rm hor}^2(R)}\right)}
  \,{\rm d} \delta_{\rm hor}(R)  =  {\rm erfc}\left(\frac{\delta_{\rm c}}{
   \sqrt{2}\sigma_{\rm hor}(R)}\right) \,. 
\end{equation}
The constraints on the initial PBH abundance are translated into
constraints on the mass variance by inverting this expression. The
constraints are scale dependent and lie in the range $\beta < 10^{-20}
- 10^{-5}$~\cite{Josan:2009qn,Carr:2009jm}. The resulting constraints
on $\sigma_{\rm hor}(R)$ lie in the range $\sigma_{\rm
  hor}(R)/\delta_{\rm c} < 0.1-0.2 $~\cite{Josan:2009qn}.

In order to calculate $\sigma_{\rm hor}(R)$ we need to evaluate the density contrast at the epoch when the scale of interest, $R^{-1} = k_{\rm c}$, enters the horizon, $c_{\rm s} k_{\rm c} = a H$, at $N=N_{k}$. We consider PBH formation for modes which re-enter the horizon after curvaton decay, $k_{\rm c} > a(N_{\rm d}) H(N_{\rm d})$, for which we can use the analytical results from section \ref{supe-sub}.

Using the Poisson equation and \eq{Ein3} the comoving density contrast is related to the Bardeen potential by
\beq
\label{dcomeq}
\frac{k^2}{a^2 H^2}\Psi = -\frac{3}{2} \delta_{\rm com}\,.
\eeq
As a result the number of PBHs formed is controlled by the amplitude of $\Psi$; the larger 
the amplitude of $\Psi$, the larger the abundance of PBH formed. As we saw before, the amplitude of $\Psi$ decreases as $f_{\rm d}$ is decreased. Let us consider the case where the PBH formation is most efficient, corresponding to $f_d \gtrsim 1/2$.
Using  Eq. (\ref{dcomeq}) and \eq{psin} for the Bardeen potential evaluated at $N=N_{k}$ with the constant $c_{1}$ given by \eq{c1-limit2}
\ba
\label{Power-sigma}
\mathcal{P}_{\delta}(k, N_k) &=& \frac{4}{9} \left(\frac{k}{ a H}\right)^4 \mathcal{P}_{\Psi}(k, N_k)  \,, \nonumber\\
&=& \frac{18 \pi}{25} \mathcal{P}_{\zeta_\sigma}  \left(\frac{k}{ a H}\right)^4
(\bar k e^{N_k})^{-3} \left[ J_{3/2} (\bar k e^{N_k}) \right]^2\,.
\ea
Using \eq{H-rad} the time of horizon crossing, $N_k$, is related to the time of curvaton decay, $N_{\rm d}$, by
\ba
e^{N_{k}} = \frac{H_{\rm d}}{k_{\rm c}} e^{2 N_{\rm d}}
\ea
so that $\bar k$, defined in \eq{kbar}, is given by
\ba
\label{bar-k-Nc}
\bar k e^{N_{k}} = \frac{k}{k_{\rm c}} \, .
\ea
Inserting the expression for the power spectrum of the comoving density contrast, \eq{Power-sigma}, 
into the definition of the mass variance, \eq{variance1}, and using \eq{bar-k-Nc} gives 
\ba
\label{Power-sigma2}
\sigma_{\rm hor}^2(R) = \frac{18 \pi}{25 c_s^4} \, \mathcal{P}_{\zeta_\sigma}
\int_0^\infty \frac{d k}{k_c} e^{-k^2/k_c^2} \left[ J_{3/2}(k/k_c) \right]^2 \,.
\ea
Finally, using the numerical approximation  
\ba
\int_0^\infty dx \,  e^{-x^2} J_{3/2}(x)^2 \simeq 0.02 \,,
\ea
we find
\ba
\label{Power-sigma3}
\sigma_{\rm hor}^2(R)    \simeq 0.4 \,  \mathcal{P}_{\zeta_\sigma} \,.
\ea
Therefore the requirement to avoid PBH overproduction, $\sigma_{\rm hor}(R)/\delta_{\rm c}  < 0.1-0.2 $,  leads to a straight-forward, and fairly weak, constraint  
$\mathcal{P}_{\zeta_\sigma}(R) < 10^{-1} - 10^{-2}$ which is easily satisfied for an almost scale invariant curvaton field.

On the other hand, if we consider the  limit in which $f_{\rm d} \ll 1$ the above result becomes $\sigma_{\rm hor}^2(R)    \sim  f_{\rm d}^2 \,  \mathcal{P}_{\zeta_\sigma}$. As a result, the condition on $\mathcal{P}_{\zeta_\sigma}(R) $ becomes even less restrictive as expected.

\section{Conclusion and discussion}

In this paper we have studied what effect the curvaton decay has on
the primordial power spectrum of the density fluctuations and the
Bardeen potential. To this end we studied a simple system comprising
only radiation and the curvaton, which we modelled as a pressureless
fluid, using a flat FRW universe as background.

The key parameter in our analysis is the weighted fraction of the
curvaton energy density to the total energy density at the time of
curvaton decay, $f_{\rm d}$, defined in \eq{fd-def}. Solving the
system of governing equations analytically, we found that on
super-horizon scales an increase in $f_{\rm d}$ will lead to an
enhancement of the amplitude of the Bardeen potential due to curvaton
dynamics. Unfortunately we were not able to solve the system in the
small scale limit analytically, and therefore leave semi-analytical
solutions in this regime to future work.

Having established an enhancement in the density contrast on
super-horizon scales, it is natural to ask whether this increase
will lead to observational consequences, in particular to the
overproduction of PBHs. We studied this issue in detail and found that
the enhancement is too small to lead to significant PBH production. We
can therefore conclude that the enhancement of the primordial power
spectrum on super-horizon scales does not lead to additional constraints
on the curvaton model through PBH production.

However, since we only found analytical solution for the
super-horizon modes, we could not investigate whether there would be
significant PBH production on sub-horizon scales. As shown in
Refs.~\cite{Lyth:2005ze} and \cite{Zaballa:2006kh}, PBH production on
sub-horizon scales can have an significant effect leading to further
constraints on the model. We hope to investigate these questions in
future work.

\section*{Acknowledgements}

We would like to thank E. Erfani and M. S. Movahed for discussions. H. F. would like
to thank Queen Mary, University of London, for hospitality where this
work was at the early stage. AMG is supported by STFC. KAM is
supported, in part, by STFC grant ST/J001546/1.  M. Z. would like to
thank ICTP for hospitality where this work was in progress.

\appendix

\section{Fluid equations in different forms}
\label{fluid-eqs}

The closed system of fluid equations in terms of variables $\{ \zeta, {\cal R}, \zeta_\sigma,  {\cal R}_\sigma$ \} is given in Eqs. (\ref{zeta-cur1})- (\ref{R1}). Here we present the equivalent systems of equations in terms of $\{ \zeta, {\cal R}, \zeta_{\rm r},  {\cal R}_{\rm r}$ \} and
$\{ \zeta_{\rm r}, {\cal R}_{\rm r}, \zeta_\sigma,  {\cal R}_\sigma$ \}.

In terms of $\{ \zeta, {\cal R}, \zeta_{\rm r},  {\cal R}_{\rm r}$ \} the system of equations read
\ba
\label{zeta-cur}
\zeta_\sigma' &=& \frac{\Gamma }{6 H }\frac{\Gamma\Omega_{\sigma}-4H\Omega_{{\rm r}}}{\Gamma+3H} S
+ \frac{k^2}{3 a^2 H^2}  \Psi - \frac{k^2}{ a^2 H^2}\left(\frac{H}{3H+\Gamma}\right){\cal R}_\sigma \,, \\
\label{zeta-gamma}
\zeta_{\rm r}' &=&- \frac{\Gamma }{6 H }\Omega_{\sigma} \frac{(3H+\Gamma)(\Omega_{\sigma}+2\Omega_{{\rm r}})}{\Gamma\Omega_{\sigma}-4H\Omega_{{\rm r}}}S  + \frac{k^2}{3 a^2 H^2} \Psi + \frac{k^2}{3 a^2 H^2}\left( \frac{4H\Omega_{{\rm r}}}{\Gamma\Omega_{\sigma}-4H\Omega_{{\rm r}}} \right) {\cal R}_{\rm r} \,,  \\
\label{R-cur}
{\cal R}_\sigma' &=& - \left( \frac{ \Gamma}{H} + \frac{H'}{H}
\right) \left(\frac{4\Omega_{{\rm r}}}{3+\Omega_{{\rm r}}}{\cal R}_{{\rm r}} +\left[\frac{3\Omega_{\sigma}}{3+\Omega_{{\rm r}}}-1\right] {\cal R}_\sigma\right)\,, \\
\label{R-gamma}
{\cal R}_{{\rm r}}' &=&  \left( \frac{3 \Gamma \Omega_\sigma}{4H
    \Omega_{\rm r} } - \frac{H'}{H} \right)
\left(\left[\frac{4\Omega_{\rm r}}{3+\Omega_{\rm r}}-1\right]
{\cal R}_{{\rm r}}
+\frac{3\Omega_{\sigma}}{3+\Omega_{{\rm r}}} {\cal R}_\sigma\right)
+ \left( 1- \frac{\Gamma \Omega_\sigma}{4H \Omega_{\rm r}}    \right) ({\cal R}_{{\rm r}} + \zeta_{{\rm r}}) \,.
\ea
Here the entropy perturbation $S$ is defined via 
\ba
\label{S}
S \equiv 3 \left( \zeta_\sigma - \zeta_{\rm r} \right)
= - 3 H \left( \frac{\delta \rho_\sigma}{\dot \rho_\sigma} -
 \frac{\delta \rho_{\rm r}}{\dot \rho_{\rm r}}
\right) \, .
\ea
We see that Eqs. (\ref{zeta-cur})-(\ref{R-gamma}) give a closed system of equations for four variables
$\{ \zeta_\alpha, {\cal R}_\alpha \}$. Note that $\Psi$ can be eliminated in these equations
from Eq. (\ref{Ein3}) in terms of $\zeta $ and ${\cal R}$ which are expressed in terms of
$\zeta_\alpha$ and ${\cal R}_\alpha$  from Eqs. (\ref{zeta-2}) and (\ref{R-2}).
As mentioned in Ref.~\cite{wands1} there is an apparent singularity in the system above when
$\dot \rho_{\rm r} =0$ and $\Gamma\Omega_{\sigma}-4H\Omega_{{\rm r}}$. In order to overcome this problem it is convenient to trade $\zeta_{\rm r}$ and ${\cal R}_{\rm r}$ for $\zeta$ and ${\cal R}$ as we did in Eqs. (\ref{zeta-cur1})- (\ref{R1}) in the main text.

Alternatively, the system of equations in terms of $\{ \zeta_{\rm r}, {\cal R}_{\rm r},
\zeta, {\cal R} \}$ is written as
\ba
\label{zeta-cur3}
\zeta_{\rm r}' &=&- \frac{\Gamma }{2 } \frac{(3 +  \Omega_{\rm r})(\Omega_{\sigma}+2\Omega_{{\rm r}})}{\Gamma\Omega_{\sigma}-4H\Omega_{{\rm r}}}(\zeta-\zeta_{{\rm r}})  + \frac{k^2}{3 a^2 H^2} \Psi + \frac{k^2}{3 a^2 H^2}\left( \frac{4H\Omega_{{\rm r}}}{\Gamma\Omega_{\sigma}-4H\Omega_{{\rm r}}} \right) {\cal R}_{\rm r}  \,, \\
\label{zeta2}
\zeta' &=& \frac{4 H \Omega_{\rm r} - \Gamma \Omega_\sigma }{H (3 +  \Omega_{\rm r})}
(\zeta - \zeta_{{\rm r}})
  + \frac{k^2}{3 a^2 H^2} (\Psi - {\cal R} )  \,, \\
  {\cal R}_{{\rm r}}' &=&  \left( \frac{3 \Gamma \Omega_\sigma}{4H \Omega_{\rm r} } - \frac{H'}{H} \right) \left({\cal R}-
{\cal R}_{{\rm r}}
\right)
+ \left( 1- \frac{\Gamma \Omega_\sigma}{4H \Omega_{\rm r}}    \right) ({\cal R}_{{\rm r}} + \zeta_{{\rm r}}) \,,
\\
\label{R3}
{\cal R}' &=&  \left( \frac{H'}{H} + \frac{4 H \Omega_{\rm r} - \Gamma \Omega_\sigma }{H (3 +  \Omega_{\rm r})} \right) {\cal R}
- \frac{1}{2}\left( 3+ \Omega_{{\rm r}} \right) \zeta + \frac{4 H \Omega_{\rm r} - \Gamma \Omega_\sigma }{H (3 +  \Omega_{\rm r})} \zeta_{\rm r} - \frac{k^2}{3 a^2 H^2}  \Psi \,.
\ea




\end{document}